\documentclass[prd,twocolumn,amsmath,amssymb,nofootinbib,preprintnumbers,balancelastpage]{revtex4}
\usepackage{graphicx}
\usepackage{hyperref}
\hypersetup{
    pdfnewwindow=true,      
    colorlinks=true,       
    linkcolor=black,          
    citecolor=blue,        
    filecolor=blue,      
    urlcolor=blue           
}
\usepackage{cancel}
\usepackage{amssymb}
\usepackage{textcomp}
\usepackage{amsmath}
\usepackage{bm}
\usepackage{times}
\usepackage{epsfig}
\usepackage{color}
\usepackage{graphics}
\usepackage{hyperref}
\usepackage{setspace}
\usepackage{slashed}
\usepackage{amsmath, amssymb}
\def\beq{\begin{equation}}
\def\eeq{\end{equation}}
\def\bea{\begin{eqnarray}}
\def\eea{\end{eqnarray}}

\begin{document}
\title{\Large  {\textit{\bf{On the Higgs Mass and Perturbativity}}}}
\bigskip
\author{\large Pavel Fileviez P{\'e}rez}
\address{Particle and Astro-Particle Physics Division \\
Max-Planck Institute for Nuclear Physics (MPIK) \\
Saupfercheckweg 1, 69117 Heidelberg, Germany}
\author{\large Sogee Spinner}
\address{International School for Advanced Studies (SISSA) \\ and INFN, Via Bonomea 265, 34136 Trieste, Italy}
\date{\today}
\begin{abstract}
The predictions for the Higgs mass in extensions of the Minimal Supersymmetric Standard Model 
are discussed. We propose a simple theory where the Higgs mass is modified 
at tree-level and one can achieve a mass around $125$ GeV without assuming heavy stops or 
large left-right mixing in the stop sector. All the parameters in the theory can be perturbative up to the grand 
unified scale, and one predicts the existence of new colored fields at the TeV scale. 
We refer to this model as ``Adjoint MSSM". We discuss the main phenomenological aspects of this scenario 
and the possible signatures at the Large Hadron Collider.
\end{abstract}
\maketitle
%
\section{Introduction}
One of the main goals of the Large Hadron Collider (LHC) is the discovery of the mechanism behind the electroweak symmetry breaking.
Recently, the ATLAS~\cite{ATLAS} and CMS~\cite{CMS} collaborations have reported the discovery of a new bosonic field with mass of about 125 GeV with dominate measured branching ratios into two photons and two massive gauge bosons: the $ZZ$ and $WW$ channels.
Currently, the ATLAS and CMS collaborations do not have enough information to conclude that this new scalar particle is responsible 
for the generation of the particle masses in the Standard Model (SM) but its properties are inline with a SM Higgs boson. 

From the theoretical point of view it is important to understand which frameworks allow a scalar field with the correct mass and properties to agree with the ATLAS and CMS data. The Minimal Supersymmetric Standard Model (MSSM) is one of the most appealing TeV scale theories and has the advantage of predicting an upper bound on the Higgs mass at tree-level. Unfortunately, the one loop corrections 
to the lightest Higgs mass in the MSSM are very sensitive to the spectrum of the stop sector. 
A Higgs mass of about 125 GeV is allowed in the MSSM but requires very heavy stops or large 
left-right mixing in the stop sector. While this is a consistent scenario in the MSSM, it is disheartening for the LHC since lighter stops and sbottoms are easier to discover.

To circumvent this constraint, the Higgs mass in extensions of the MSSM have been investigated by many groups:
see the reviews in Ref.~\cite{Carena} and Ref.~\cite{Ellwanger:2009dp} for details and references therein. 
One of the most studied models is the Next-to-Minimal Supersymmetric Standard Model (NMSSM) where the Higgs sector is composed of the MSSM Higgs fields 
and an extra singlet which can lift the tree-level prediction above the MSSM value. The singlet-Higgs coupling generates an extra contribution to the Higgs mass at 
tree-level, which can, in principle, be large enough to modify the Higgs mass to 125 GeV even without contribution from the quantum corrections. However, for this to happen the coupling between the MSSM Higgs fields and the new singlet must be so large that it becomes non-perturbative far below the GUT scale and therefore the NMSSM would requires a low cutoff. Since gauge coupling unification at the high scale is a desirable feature of the MSSM (and NMSSM) particle content and since one of the appealing features of SUSY itself is its calculability up to the GUT scale, it is imperative to search for a theory that can do both: raise the tree-level Higgs mass and remain perturbative into the ultraviolet regime. This is the main goal of this paper.

In this work we propose a simple theory where the lightest Higgs mass can be large enough and all relevant parameters can be perturbative up to the unified scale with light stops and little stop mixing. We refer to this model as the ``Adjoint MSSM'' as the MSSM particle content is extended by an adjoint ($24$) representation of $SU(5)$. In order to solve the $\mu$-problem, a $Z_3$ symmetry is imposed so that all masses result from the vacuum expectation value (VEV) of the singlet in the $24$, as in the NMSSM. We investigate the predictions of the tree-level Higgs mass and the perturbativity of the couplings in this theory and compare to some MSSM extensions such as the NMSSM and the triplet extended NMSSM. As our benchmark we seek parts of parameter space where the Higgs mass is about 110 GeV at tree-level, so that 500 GeV stops with little mixing can lift it to 125 GeV, and that all parameters are perturbative up to the GUT scale. Since all masses arise from the singlet VEV the colored fields in the $24$ are at the TeV scale which help to keep the perturbative behavior of the Yukawa couplings and the maintain the unification of the gauge couplings. Such an effect of extra matter was also previously discussed in a different extension of the NMSSM in~\cite{Masip:1998jc}. We also briefly discuss the possible decays and signatures of these additional fields at the Large Hadron Collider.

This article is organized as follows: In Section II we discuss the predictions for the Higgs mass at tree-level in the NMSSM, 
while in Section III we discuss an extended Higgs sector where one has a real $SU(2)_L$ triplet. In Section IV we propose 
the ``Adjoint MSSM" model and discuss the predictions for the Higgs mass. In the appendices we show the explicit form of the 
mass matrices of the scalars and fermions in the theory. 
\section{Higgs Mass in the NMSSM}
It is well-known that the light Higgs boson mass in the MSSM is given by
\begin{equation}
m_h^2 = M_Z^2 \  \cos^2 2 \beta \ + \  \delta m_h^2, 
\end{equation}
where $M_Z$ is the mass of the Z boson and the angle $\beta$ is defined by the ratio of the expectation values of the two Higgses in the theory, $\tan \beta=v_u / v_d$.
In order to achieve a mass of about 125 GeV when the tree-level mass is close to $M_Z$ (large $\tan \beta$), the one-loop correction $\delta m_h^2$ has to be around $86$ GeV. 
In the context of the MSSM it is possible only if the left-right mixing in the stop sector, $X_t$, is large when the stops are light, i.e. below the TeV scale. 
See for example the study in Ref.~\cite{Arbey:2011ab}. Even if it is possible to have a scenario with large $X_t$, it is interesting to investigate the possibility of achieving a large Higgs mass without a large $X_t$ values.

One possibility is to consider the NMSSM where a new Higgs singlet field is added to the MSSM and a $Z_3$ symmetry imposed to avoid the so-called $\mu$-problem. The effective $\mu$-term coupling is also the coupling that generates an extra contribution to the Higgs mass at tree-level. The NMSSM has been discussed by many groups and has been reviewed in Ref.~\cite{Ellwanger:2009dp}. 
See also the discussion in Ref.~\cite{Barbieri:2011tw}.

In the NMSSM the Higgs superpotential reads as
\begin{equation}
	{\cal W}_S = - \lambda_H  \hat{S}  \hat{H}_u^T i \sigma_2 \hat{H}_d \ + \ \frac{1}{3} \kappa \hat{S}^3.
\end{equation}
Here $\hat{H}_u \sim (1,2,1/2)$, and $\hat{H}_d \sim (1,2,-1/2)$ are the Higgs chiral superfields 
and $\hat{S} \sim (1,1,0)$ is the extra singlet. The relevant soft SUSY breaking Lagrangian is given by
\begin{eqnarray}
	-\mathcal{L}_{\rm{soft}}^{(1)} &=& \ m_{H_d}^2 \left| H_d \right|^2  + m_{H_u}^2 \left| H_u \right|^2 +
	m_{S}^2 \left| S \right|^2 \nonumber \\
	&+& \left(
		-  a_H S H_u^T i \sigma_2 H_d + \frac{1}{3} a_\kappa S^3 + \text{h.c}
	\right).
\end{eqnarray}
In this case the upper bound on the Higgs mass is modified at tree-level to 
\begin{equation}
	\label{nmssm.upper.bound}
	m_h^2 = M_Z^2 \  \cos^2 2 \beta \ + \   \frac{1}{2} \lambda_H^2 v^2 \sin^2 2 \beta,
\end{equation}
where $v^2=v_u^2 + v_d^2=(246 \ \rm{GeV})^2$. Therefore, if $\tan \beta$ is small the new contribution can become large (at the price of the lowering the MSSM contribution). 
Regardless, a higher tree-level mass can be achieved in the NMSSM. However a Higgs mass of 125 GeV cannot be achieved without heavy stops and/or large stop 
mixings or non-perturbative $\lambda_H$ values below the GUT scale, $M_{GUT} \approx 10^{16}$ GeV. Therefore, in this case it is not clear if one can keep the 
unification of the gauge couplings at the GUT scale because the cutoff of the theory is much lower, see for example Ref.~\cite{Barbieri:2011tw}.
Furthermore, there are several open questions in the context of the NMSSM, \textit{e.g.} 
What is the origin of the Higgs singlet and its interactions?  Is it possible to keep the perturbative 
behavior of $\lambda_H$ in a simple extension of the NMSSM? In this paper we will present a such simple extension 
of the MSSM where we can keep the perturbative behavior of all couplings up the high scale. 
\begin{figure}[h] 
	\includegraphics[scale=0.7]{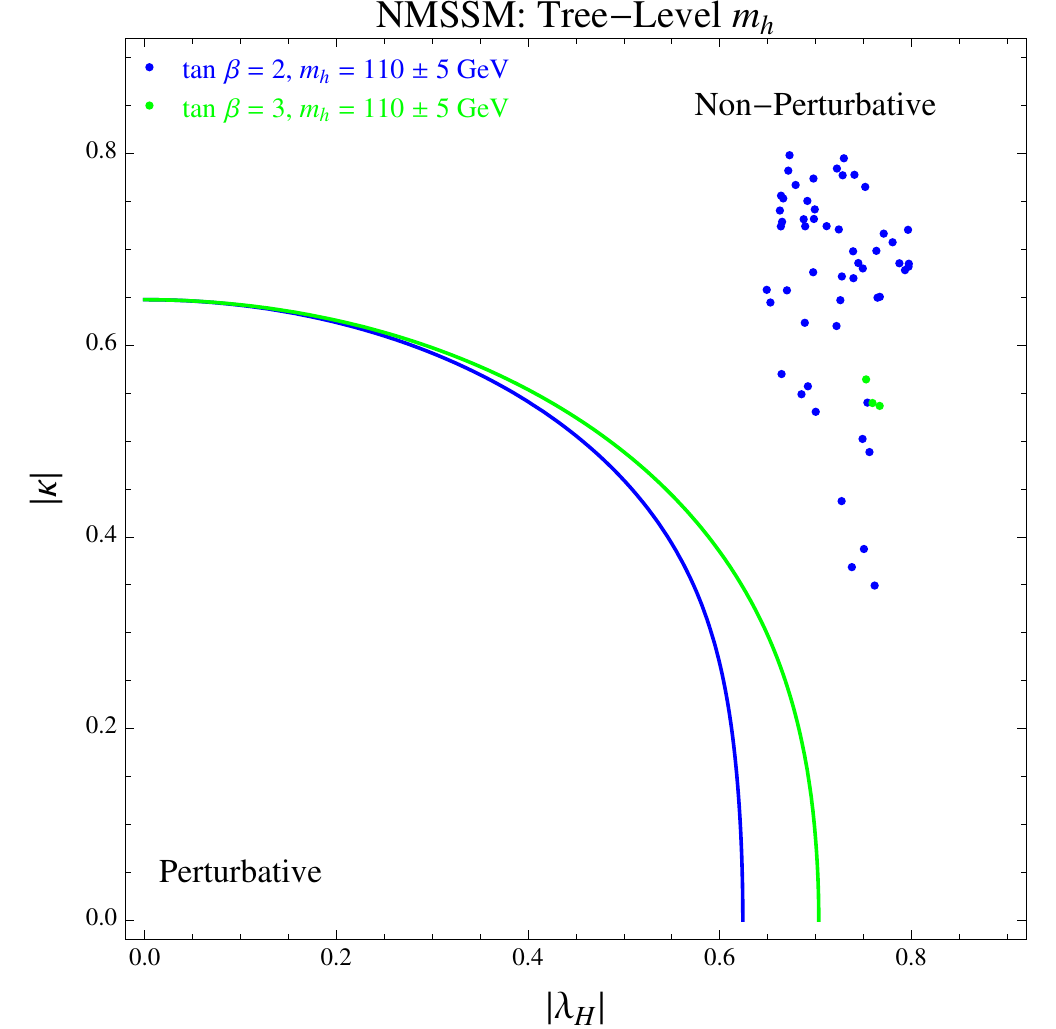}
	\caption{Here we show curves that separate the perturbative region (below and to the left of the curves) from the non-perturbative region (above and to the right of the curves) in the $\kappa-\lambda_H$ plane in the NMSSM. Blue and green curves and points correspond to $\tan \beta = 2$ and $\tan \beta = 3$, respectively. The points represent a Higgs mass of $110 \pm 5 $ GeV resulting from a scan of parameters as in the text. This approximately corresponds to the necessary tree-level Higgs mass with stops at 500 GeV and no mixing. From this plot, one can see that it is not possible to achieve a 110 GeV Higgs mass at tree-level that is consistent with the perturbative bounds on the system in the NMSSM.}
	\label{img.pert.mh110.nmssm}
\end{figure}
\begin{figure}[h!] 
	\includegraphics[scale=0.7]{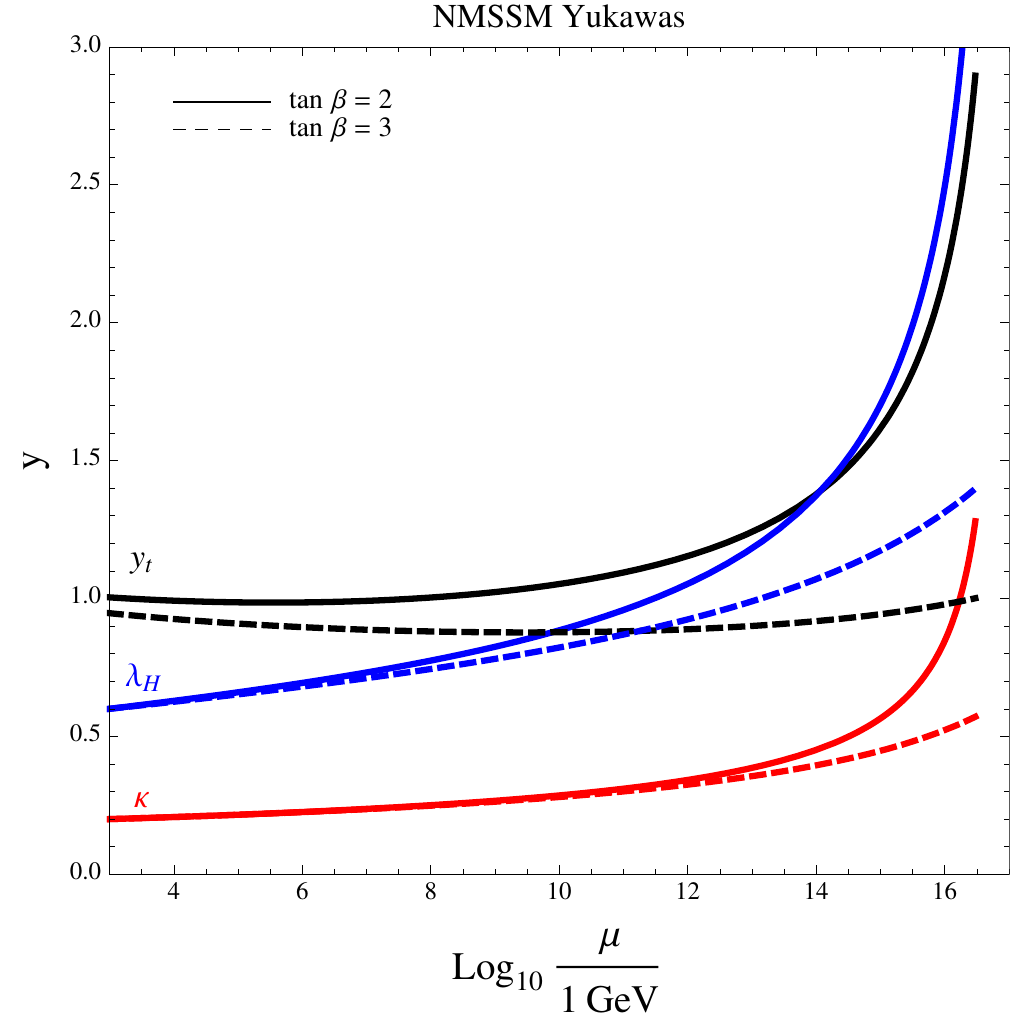}
\caption{Evolution of the couplings in the NMSSM where solid (dashed) curves represent $\tan \beta =2$ ($\tan \beta =3$) and $\lambda_H = 0.6$ at the low scale.}
\label{img.NMSSM.y}
\end{figure}
Before doing so, we first review the case of the NMSSM. Our guidelines are as follows:
\begin{itemize}
	\item A tree-level Higgs mass of $110 \pm 5$ GeV (this requires 500 GeV stops and little stop mixing to raise it to 125 GeV).
	\item All couplings perturbative up to the GUT scale, which we take to be about $2 \times 10^{16}$ GeV.
\end{itemize}
Fig.~\ref{img.pert.mh110.nmssm} measures the NMSSM performance under these guidelines in the $\kappa-\lambda_H$. The curves separate the perturbative regime (below and to the left of a given curve) from the non-perturbative regime (above and to the right of a given curve).The blue and green curves and points correspond to $\tan \beta = 2$ and $\tan \beta = 3$, respectively. The points represent a Higgs mass of $110 \pm 5$ GeV resulting from a scan of the following parameters:
\begin{center}
$0.1 \leq | \lambda_{H} |, |\kappa|  \leq 0.7$, \quad  $100 \  \rm{GeV}  \leq  |\mu|  \leq 1 \  \rm{TeV} $, \quad  $0 \leq |a_{H}| , |a_{\kappa}| \leq 1 $ TeV,
\end{center}
where $\mu= \lambda_H v_S / \sqrt{2}$. In addition to requiring the appropriate mass, we also require that the point in parameter space correspond to the decoupling limit (the mostly MSSM pseudoscalar has mass greater than 300 GeV), the mostly singlet pseudoscalar be heavier than half the Higgs mass (125 GeV) and that in cases where the lightest CP-even scalar is mostly singlet that its mass be larger than 90 GeV. The former two conditions are imposed so that the Higgs branching ratios are inline with a SM-like Higgs while the latter is a conservative bound on a light mostly singlet scalar to keep it consistent with LEP 2, however they do not change the general outcome of this study. Namely, from Fig.~\ref{img.pert.mh110.nmssm} it is clear that the NMSSM does not adhere to our guidelines: $110 \pm 5$ GeV tree-level mass and all couplings perturbative to the GUT scale.

An important lesson here is the role of $\tan \beta$, which can already be observed from Fig.~\ref{img.pert.mh110.nmssm}: an increase in $\tan \beta$ worsens the perturbativity of the NMSSM at the same time as raising the Higgs mass as indicated in Eq.~\ref{nmssm.upper.bound}. To illustrate this point further, the evolution of the relevant couplings (the top Yukawa coupling, $\kappa$ and $\lambda_H$) are shown in Fig.~\ref{img.NMSSM.y} where solid (dashed) curves represent $\tan \beta =2$ ($\tan \beta =3$) and $\lambda_H = 0.6$ at the TeV scale. This value of $\lambda_H$ sits at the edge of the perturbative regime and the two values of $\tan \beta$ highlight the fact that it is the small increase in the top Yukawa coupling due to the decrease of $\tan \beta$ that starts to spoil perturbativity. Therefore, additional particle content, which would increase the negative contributions in the top Yukawa RGE, Eq.~\ref{beta.y.top}, might make it possible to use lower values of $\tan \beta$ while keeping the system perturbative as was done for example in~\cite{Agashe:2012zq}. Regardless, since we cannot achieve solutions in the perturbative region it is important to go beyond the NMSSM to see if one can find solutions without assuming a low cutoff.
\section{The Triplet Extended NMSSM and the Higgs Mass}
Possible modifications of the Higgs mass at tree-level can be understood by considering all of the possible 
fields which can couple to $H_u$ and $H_d$. Above we have discussed the well-known impact of the singlet 
field in the NMSSM, but one can also consider the impact of a triplet with zero hypercharge as well.
For previous studies of supersymmetric models with real triplets see Refs.~\cite{Espinosa:1991wt, 
Espinosa:1991gr, DiChiara:2008rg, Basak:2012bd, Delgado:2012sm}.
Since each type of field leads to a new contribution to the tree-level Higgs mass, it is logical that the presence of both types of fields will yield more favorable results. Therefore, in this section we study the Higgs sector composed of a singlet, a triplet with zero hypercharge 
and the Higgs doublets, the triplet extended NMSSM (TNMSSM)~\cite{Basak:2012bd}. In this case the relevant superpotential can be written as
\begin{eqnarray}
	{\cal W}_{S\Sigma} &=& - \lambda_H  \hat{S}  \hat{H}_u^T i \sigma_2 \hat{H}_d \ + \ \frac{1}{3} \kappa \hat{S}^3 \ + \
	\frac{1}{2} \lambda_\Sigma \, \hat{S} \, \text{Tr} \, \hat{\Sigma}^2 \nonumber \\
	& - &  \lambda_{3}  \hat{H}_u^T i \sigma_2  \hat{\Sigma}  \hat{H}_d.
\end{eqnarray}
Here the $\hat{\Sigma} \sim (1,3,0)$ field is given by
\begin{align}
\begin{split}
	\hat{\Sigma} =
	\begin{pmatrix}
		\frac{1}{\sqrt 2} \hat{\Sigma}^0
		&
		\hat{\Sigma}^+
		\\
		\hat{\Sigma}^-
		&
		- \frac{1}{\sqrt 2} \hat{\Sigma}^0
	\end{pmatrix}.
\end{split}
\end{align}
Notice that the $\rm{Tr} \Sigma=\rm{Tr} \Sigma^3=0$ and that a $Z_3$ symmetry has been enforced so that no $\mu$-problem exists. The soft SUSY breaking Lagrangian relevant for the Higgs sector is given by
\begin{eqnarray}
	-\mathcal{L}_{\rm{soft}}^{(2)} & = & - \mathcal{L}_{\rm{soft}}^{(1)}  \ + \ 
	m_\Sigma^2  \text{Tr} \Sigma^2 \nonumber \\
	&+ &\left( \frac{1}{2} a_\Sigma \, S \, \text{Tr} \, \Sigma^2 -
		a_3 H_u^T i \sigma_2 \Sigma H_d \ + \  \text{h.c} \right).
\end{eqnarray}
In this case the upper bound of the lightest Higgs mass at tree-level can be lifted to
\begin{equation}
	m_h^2 = M_Z^2 \  \cos^2 2 \beta \ + \   \frac{1}{2}\lambda_H^2 v^2 \sin^2 2 \beta  \ + \   \frac{1}{4}\lambda_3^2 v^2 \sin^2 2 \beta.
\end{equation}
Therefore, if $\tan \beta$ is small with $\lambda_H$ and $\lambda_3$ large one can achieve a Higgs mass even larger than the one in the NMSSM. However, it is not clear if it is possible to satisfy the perturbative bounds or if one needs to assume a low cutoff. Since this model is not as well known as the NMSSM, before tackling this question, we present some more details of the TNMSSM.
As in any SUSY model, the breaking of the electroweak symmetry in this theory results from the soft terms above, the contribution from the F-terms and the Higgs potential from the $D$-terms. For the latter, it is necessary to state the kinetic term for the triplet, which can be written as
\begin{equation}
{\cal L}_{Kin}^{\Sigma} = \int d^4 \theta \  \rm{Tr} \ \hat{\Sigma}^\dagger \left[ e^{2 g_2 \hat{V}_2} , \hat{\Sigma} \right] .
\end{equation}
Using this expression one can find the D-term contribution to the scalar potential:
\begin{eqnarray}
	V_D &=& \frac{1}{8} g_1^2 \left( \left| H_u \right|^2 - \left| H_d \right|^2 \right)^2 + \nonumber \\
	&& \frac{1}{8} g_2^2 \left(H_u^\dagger \sigma_a H_u + H_d^\dagger \sigma_a H_d + 2 \, \text{Tr} \, \Sigma^\dagger \sigma_a \Sigma\right)^2
\end{eqnarray}
summed over $a$. The VEVs are defined as follows:
\begin{equation}
	\left<H_{u}\right> = \frac{v_u}{\sqrt 2},
	\quad 
	\left<H_{d}\right> = \frac{v_d}{\sqrt 2},
	\quad 
	\left<S\right> = \frac{v_S}{\sqrt 2},
	\quad 
	\left<\Sigma \right> = \frac{v_\Sigma}{\sqrt 2}.
\end{equation}
Since the VEV of the triplet breaks custodial symmetry it cannot be large, \textit{i.e.} $v_\Sigma \lesssim 4$ GeV, 
see for example Ref.~\cite{FileviezPerez:2008bj} for a recent discussion. 
The minimization conditions, neglecting the terms proportional to $v_\Sigma$, 
are given by
\begin{widetext}
\begin{eqnarray}
	m_{H_u}^2  - \frac{1}{2} M_Z^2 c_{2\beta}
	+\frac{1}{2} \lambda_H^2 v_S^2+ \frac{1}{2}c_\beta^2 \lambda_H^2 v^2
	+ \frac{1}{4} \lambda_3^2 c_\beta^2 v^2
	+\frac{v_S}{t_\beta}
	\left(
		\frac{1}{\sqrt 2} a_H + \frac{1}{2} \lambda_H \kappa v_s
	\right) & = & 0,
	\\
	m_{H_d}^2 +\frac{1}{2} M_Z^2 c_{2\beta}  + \frac{1}{2}\lambda^2 v_S^2
	+ \frac{1}{4} s_\beta^2 \lambda_3^2 v^2 + \frac{1}{2} s_\beta^2 \lambda_H^2 v^2 
	+t_\beta v_S
	\left(
		\frac{1}{\sqrt 2} a_H + \frac{1}{2} \lambda_H \kappa v_S
	\right) &=& 0,
\\
m_S^2 + \kappa^2 v_S^2 + \frac{1}{2} \lambda^2 v^2 + \frac{1}{\sqrt 2} a_\kappa v_S
	+\frac{1}{2} s_{2 \beta} v^2
	\left(
		\frac{1}{\sqrt 2} \frac{a_H}{v_S} + \lambda_H \kappa
	\right) &=& 0,
	\\
	m_\Sigma^2 + \frac{1}{2} \lambda_\Sigma^2 v_S^2 + \frac{1}{4} \lambda_3^2 v^2
	 + \frac{1}{2} \lambda_\Sigma \kappa v_S^2 + \frac{1}{\sqrt 2} a_\Sigma v_S
	+\frac{1}{4} s_{2 \beta} \lambda_H \lambda_\Sigma v^2
	&&
	\nonumber
	\\
	+ \frac{ v^2}{2 \sqrt 2 \, v_\Sigma}
	\left[
		\frac{1}{\sqrt 2} s_{2 \beta}a_3 + \frac{1}{2} s_{2 \beta} \lambda_3 \lambda_\Sigma v_S  +\lambda_H \lambda_3 v_S 
	\right] 
	&=& 0,
\end{eqnarray}
\end{widetext}
where $c_x, s_x, t_x$ represents the $\cos x, \sin x$ and $\tan x$ respectively. The VEV of $\Sigma^0$ can be approximated as
\begin{equation}
	v_\Sigma \approx
	-\frac{v^2}{4 \sqrt 2}
	\frac
	{\left( 	\sqrt 2 s_{2 \beta}a_3
		\ + \  \lambda_3 v_S \left( s_{2 \beta} \lambda_\Sigma  +\lambda_H \right) \right)}{M_\Sigma^2 },
\end{equation}
where
\begin{equation}
M^2_{\Sigma}= m_\Sigma^2 + \frac{1}{2} v_S^2
		\left(
			\lambda_\Sigma^2 + \lambda_\Sigma \kappa
		\right)
		+ \frac{1}{\sqrt 2} a_\Sigma v_S + \frac{1}{4} v^2
		\left(
			\lambda_3^2 + s_{2 \beta} \lambda_H \lambda_\Sigma
		\right).
\end{equation}
Note that there exists some tension between increasing the Higgs mass through $\lambda_3$ and keeping the triplet VEV small.
\begin{figure}[ht] 
	\includegraphics[scale=0.7]{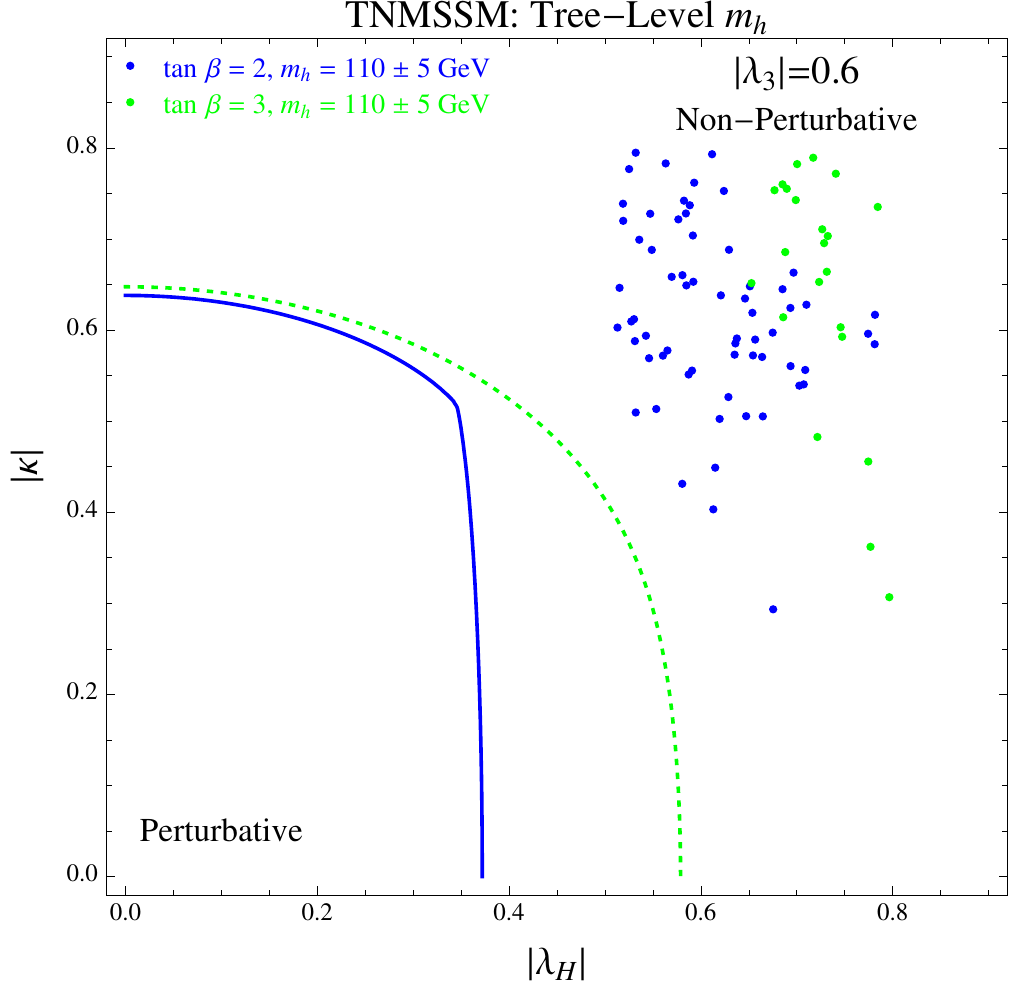}
	\caption{We show the curves that separate the perturbative region (below and to the left of the curves) from the non-perturbative region (above and to the right of the curves) in the $\kappa-\lambda_H$ plane in the TNMSSM for we have used $\lambda_3 = 0.6$ and $\lambda_\Sigma=0.2$. Blue and green 
	lines and points correspond to $\tan \beta = 2$, and $\tan \beta=3$, respectively. The points represent a Higgs 
	mass of $110 \pm 5 $ GeV resulting from a scan of parameters as in the text. This approximately corresponds 
	to the necessary tree-level Higgs mass with stops at 500 GeV and no mixing.}
\label{img.pert.mh110.tnmssm}
\end{figure}
In Fig.~\ref{img.pert.mh110.tnmssm} we investigate the TNMSSM with $|\lambda_3| =0.6$ given our earlier guidelines of a tree-level Higgs mass of $110 \pm 5$ GeV and couplings perturbative to the GUT scale (even though gauge coupling unification is no longer possible), as was done for the NMSSM in Fig.~\ref{img.pert.mh110.nmssm}. The curves separate the perturbative regime (below and to the left of the curves) 
from the non-perturbative regime (above and to the right of the curves) in the $\kappa-\lambda_H$ plane. Blue (green) represents $\tan \beta = 2$ ($\tan \beta = 3$) for both the curves and the points. The points indicate a tree-level 
Higgs mass of $110 \pm 5$ GeV resulting from a scan of the TNMSSM parameters:
\begin{eqnarray}
	\label{scan}
 	&& 0.1 \  \leq \  | \lambda_{H} |,  | \lambda_{\Sigma} |,  | \kappa |  \   \leq  0.7,
	 \quad 100 \  \rm{GeV} \  \leq \ |\mu| \  \leq 1 \text{ TeV}, \nonumber \\
	&& \quad 0 \  \leq  \  |a_{\Sigma}|,  |a_{H}|,  |a_{3}|, | a_{\kappa}|  \  \leq  \ 1 \text{ TeV}, 
\end{eqnarray}
and we solve for $m_{\Sigma}$ obtaining $340 \  \rm{GeV} \  \leq \ m_\Sigma \  \leq \ 6450$ GeV given $|v_\Sigma| = 1$ GeV.
We have applied the same constraints as in the NMSSM analysis and as before this Higgs mass value roughly correspond to attaining a 125 GeV Higgs, in line with the recent LHC discovery, given 500 GeV stops with small left-right mixing in the stop sector. 

\begin{figure}[t] 
	\includegraphics[scale=0.7]{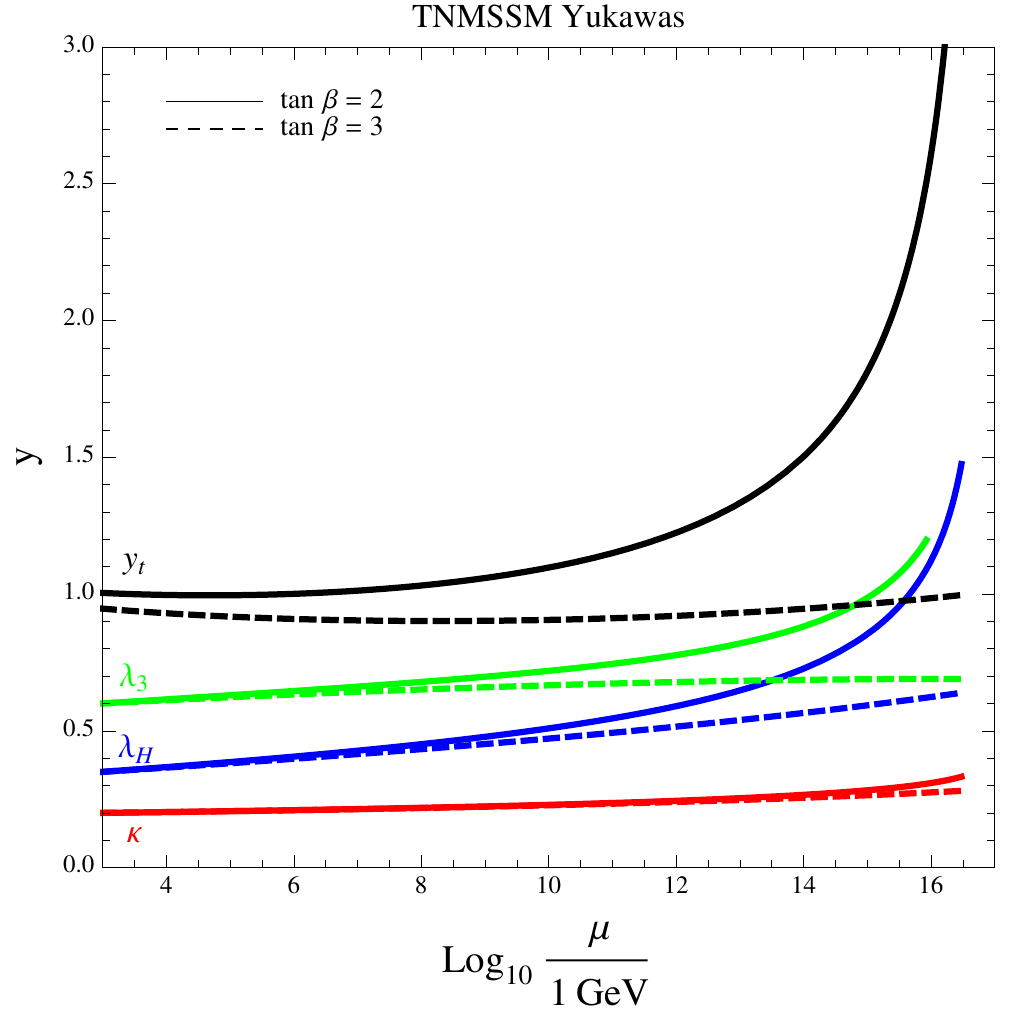}
	\caption{Evolution of the couplings in the TNMSSM where solid (dashed) curves represent $\tan \beta =2$ ($\tan \beta =3$).}
\label{img.TNSSM.y}
\end{figure}

A comparison between Figs.~\ref{img.pert.mh110.nmssm} and~\ref{img.pert.mh110.tnmssm} shows that while the lower values of $\lambda_H$ are now necessary 
for a 110 GeV tree-level Higgs mass, the perturbative behavior has become worse and its decline with decreased $\tan \beta$ has accelerated. We have investigated 
other values of $\lambda_3$ and they do not improve the behavior. Therefore, the TNMSSM does not fit our guidelines of perturbativity and large enough Higgs mass.
The running of the Yukawa couplings in the TNMSSM are displayed in Fig.~\ref{img.TNSSM.y} where solid (dashed) curves represent $\tan \beta =2$ ($\tan \beta =3$). The same lesson that was evident in the NMSSM is more dramatic here: while an $\tan \beta$ close to one is advantageous for the Higgs mass, it is quite detrimental for perturbativity.
Also  of note is gauge couplings unification, as in the MSSM, is no longer possible because the real triplet modifies
the running of $g_2$ in a significant way. We therefore conclude that this modification is not very appealing.
In the next section we go beyond the TNMSSM and apply our guidelines to the adjoint MSSM testing the tree-level Higgs mass 
and the perturbativity of the couplings.
\section{The Adjoint MSSM}
We have discussed in the previous section the impact of the singlet and the triplet fields on the Higgs mass which can be quite large. 
However, in this case gauge couplings no longer unify due to the impact of the triplet and 
achieving a large enough Higgs mass is not consistent with all parameters perturbative up to the GUT scale. 
Then, what is the next step?. In order to preserve the unification of gauge couplings 
of the MSSM we need to add extra multiplets to have a complete representation of $SU(5)$. 
In this case we will have the following extra fields
\begin{equation}
	\Phi \sim \left(8,1,0 \right), \ 
	X \sim (3, 2, - 5/6), \  \  \rm{and} \  \ 
	\bar X \sim (\bar 3, 2, 5/6),
\end{equation}
which when added to the singlet and triplet form a complete $24$ representation. 
Therefore, we refer to this model as the ``Adjoint MSSM". In this case the superpotential reads as
\begin{equation}
	{\cal W}_\text{Adj} = {\cal W}_{Y} \ + \ {\cal W}_{S\Sigma} \ + \  {\cal W}_{C} \ + \  {\cal W}_{B},
\end{equation}
where in ${\cal W}_{Y}$ one has the Yukawa interactions of the quarks and leptons 
\begin{equation}
{\cal W}_{Y} = Y_u \hat{Q} \hat{H}_u \hat{u}^c \ - \  Y_d \hat{Q} \hat{H}_d \hat{d}^c \ - \  Y_e \hat{L} \hat{H}_d \hat{e}^c, 
\end{equation}
and the interactions of the colored fields are in
\begin{eqnarray}
	{\cal W}_C &=& \frac{1}{3} \eta \text{Tr} \, \hat{\Phi}^3
	\ + \ \zeta_\Phi \hat {X} \hat \Phi \hat {\bar X} \ + \ \zeta_\Sigma  \hat X \hat \Sigma \hat {\bar X} \nonumber \\
	& + &  \frac{1}{2} \lambda_\Phi \hat{S} \, \text{Tr} \, \hat{\Phi}^2 + \lambda_X \hat{S} \, \hat{X}  \hat{\bar X},
\end{eqnarray}
where the last two terms serve as mass terms for the colored fields once $S$ acquires a VEV. In order to 
allow the $X$ field to decay we need to add the interaction between $X$, the Higgs and the SM down 
quark which is allowed by the symmetry. Therefore, one has
\begin{equation}
	{\cal W}_{B} =  y_X  \hat{X}  \hat{H}_u \hat{d}^c \ + \  \lambda^{''}  \hat{u}^c  \hat{d}^c \hat{d}^c.
\end{equation}
We have included the last term in the above equation because in principle if the $X$ field does not have 
baryon number the first term breaks baryon number at the renormalizable level. Notice again 
that the superpotential is invariant under the discrete symmetry, $Z_3$, where all superfields transform 
as $\Psi \  \to \  \exp{i 2\pi/3} \ \Psi$, and therefore there is no $\mu$ problem since all fields get mass after symmetry breaking.
Here we do not include the couplings which break lepton number, since one could assume a leptonic discrete symmetry 
similar to  R-parity.
\begin{figure}[h] 
	\includegraphics[scale=0.7]{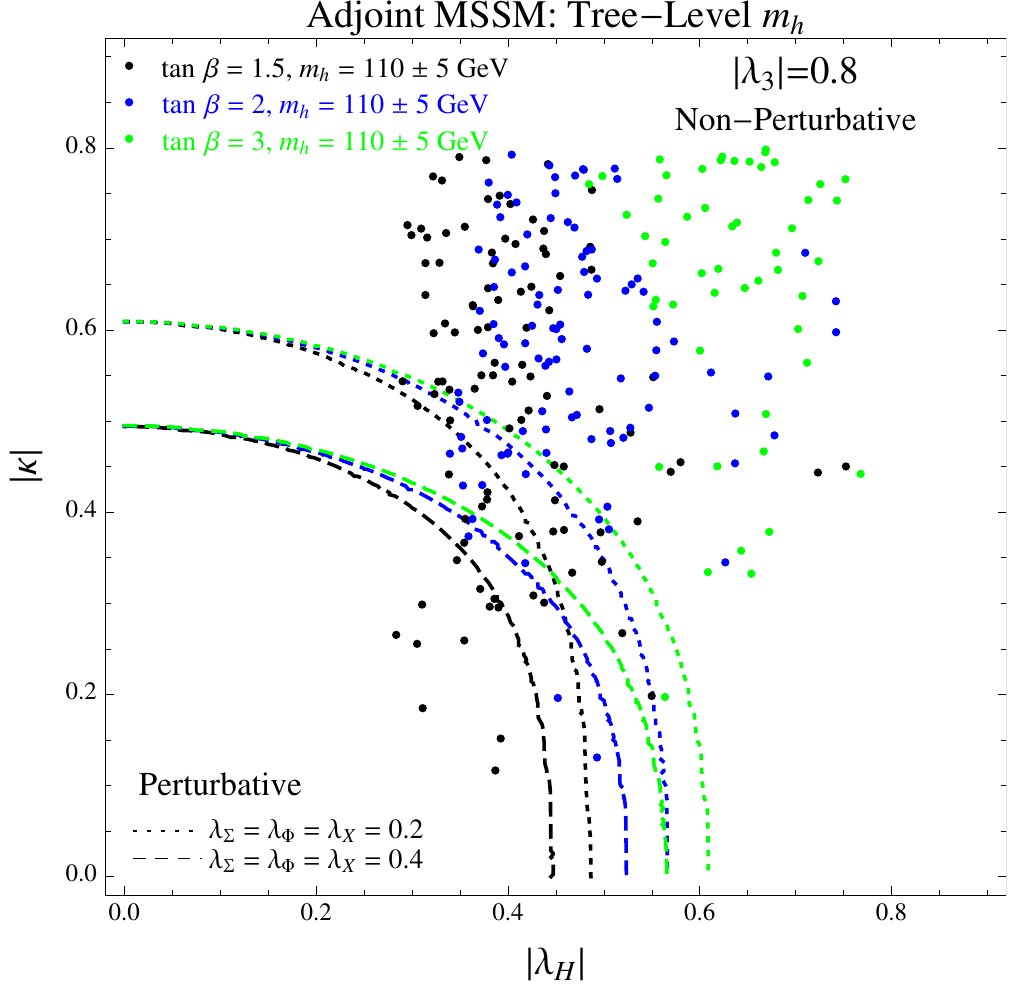}
	\caption{Presented here are the perturbative regions (below and to the left of the curve) and the non-perturbative regions (above and to the right) for $\lambda_3 = 0.8$ in the $\kappa-\lambda_H$ plane. Dotted (dashed)  curves correspond to $\lambda_\Sigma = \lambda_\Phi = \lambda_X = 0.2$ ($\lambda_\Sigma = \lambda_\Phi = \lambda_X = 0.4$) while black, blue and green lines and points correspond to $\tan \beta = 1.5, 2, 3$ respectively. The points represent a Higgs mass of $110 \pm 5 $ GeV and a results of a scan as shown in the text. This approximately corresponds to the necessary tree-level Higgs mass with stops at 500 GeV and no mixing. From this plot, one can determine if for certain values of $\lambda_\Sigma, \lambda_\Phi, \lambda_X$ and $\tan \beta$ it is possible to get a 110 GeV Higgs at tree-level that is consistent with the perturbative bounds on the system. The new Yukawa couplings not mentioned in the plot have been chosen as $y_X=\eta=\zeta_\Phi=\zeta_\Sigma=0.05$.}
\label{img.pert.mh110.l30p8}
\end{figure}
\begin{figure}[h] 
	\includegraphics[scale=0.7]{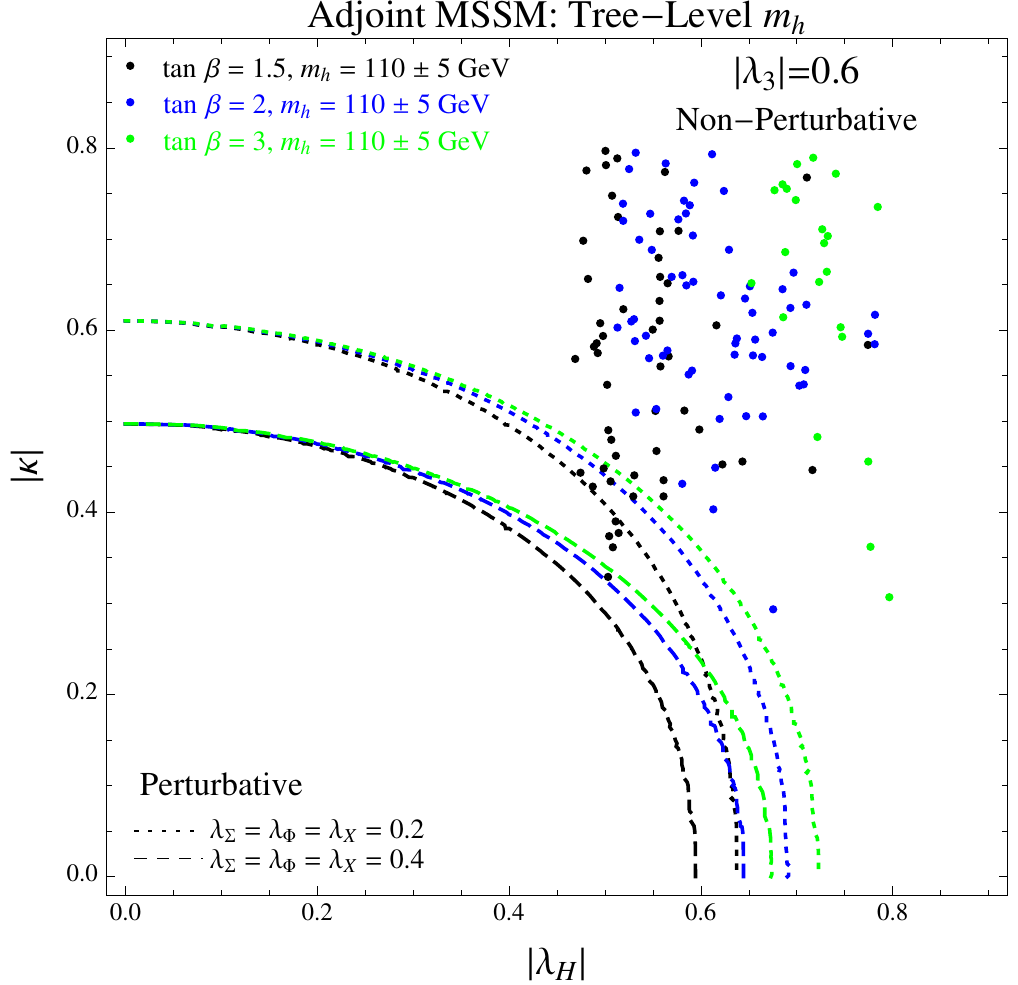}
	\caption{Presented here are curves that separate the perturbative regions (below and to the left of the curve) from the non-perturbative region (above and to the right) for $\lambda_3 = 0.6$ in the $\kappa-\lambda_H$ plane. Dotted (dashed)  curves correspond to $\lambda_\Sigma = \lambda_\Phi = \lambda_X = 0.2$ ($\lambda_\Sigma = \lambda_\Phi = \lambda_X = 0.4$) while black, blue and green lines and points correspond to $\tan \beta = 1.5, 2, 3$ respectively. The points represent a Higgs mass of $110 \pm 5 $ GeV and a results of a scan as shown in the text. This approximately corresponds to the necessary tree-level Higgs mass with stops at 500 GeV and no mixing. The new Yukawa couplings not mentioned in the plot have been chosen as $y_X=\eta=\zeta_\Phi=\zeta_\Sigma=0.05$.}
\label{img.pert.mh110.l30p6}
\end{figure}

We proceed by making a similar study of the adjoint MSSM as was done in the NMSSM and TNMSSM, namely applying the guidelines of 110 GeV Higgs 
mass and perturbative couplings. Therefore, in Fig.~\ref{img.pert.mh110.l30p8} and Fig.~\ref{img.pert.mh110.l30p6} we show curves that separate the 
perturbative regions (below and to the left of the curve) from the non-perturbative region (above and to the right) for $\lambda_3 = 0.8$ and $\lambda_3 = 0.6$ in the 
$\kappa-\lambda_H$ plane. Dotted (dashed) curves correspond to $\lambda_\Sigma = \lambda_\Phi = \lambda_X = 0.2$ 
($\lambda_\Sigma = \lambda_\Phi = \lambda_X = 0.4$). Black, blue and green lines and points correspond 
to $\tan \beta = 1.5, 2, 3$, respectively. As in the previous studies, the points represent a Higgs mass of $110 \pm 5 $ GeV 
and are the results of a scan over the rest of parameter space as in Eq.~\ref{scan}. The new Yukawa couplings not mentioned in the plot have been chosen as $y_X=\eta=\zeta_\Phi=\zeta_\Sigma=0.05$.

Some comments are now in order. As mentioned above, extra particle content can help to keep the running of the Yukawa couplings under more control since it increases the value of the gauge couplings, which in turn reduce the value of the Yukawa couplings. This role is played by the colored fields and real triplet of the $SU(5)$ adjoint. Specifically, in this case values of $\tan \beta = 1.5$ are still allowed even though they were not possible in the previous studies. It is furthermore clear that the decreased value of $\tan \beta$ value gives a significant boost to the Higgs mass.

The couplings $\lambda_\Sigma, \lambda_X$ and $\lambda_\Phi$ control the mass of the real triplet and colored fermions respectively. Obviously, they also play a major role in the perturbativity of the Yukawa couplings. To get a sense of the masses that correspond to the couplings plotted in Figs.~\ref{img.pert.mh110.l30p8} and~\ref{img.pert.mh110.l30p6} we note that points plotted correspond to $|v_S|$ in the range of 400 to 4000 GeV, translating into mass in the range 
of 57 GeV - 570 GeV (113 GeV - 1130 GeV) for $\lambda_\Sigma = \lambda_X = \lambda_\Phi = 0.2$ ($\lambda_\Sigma = \lambda_X = \lambda_\Phi = 0.4$). While the lower end of these ranges are not realistic, phenomenological points that meet our criteria still exist. As for the remaining Yukawa couplings ($\eta, y_X, \zeta_\Sigma$ and $\zeta_\Phi$) we have chosen them insignificant with respect to the RGE evolution although, of course, if they were larger they would effect the perturbativity of the couplings. Some of these couplings control the decays of the colored fields however their lower bounds are significantly below the values chosen for the plots.  We will discuss the phenomenology and bounds on these fields in a later section

Comparing Figs.~\ref{img.pert.mh110.l30p8} and~\ref{img.pert.mh110.l30p6} we see that in addition to $\tan \beta$, $\lambda_3$ also plays an important role in meeting the guidelines that we have outlined. Regardless we have come across a model that meets our criteria of a tree-level $110 \pm 5$ GeV Higgs mass with all couplings perturbative to the GUT scale: 
The Adjoint MSSM.

\begin{figure}[h] 
	\includegraphics[scale=0.7]{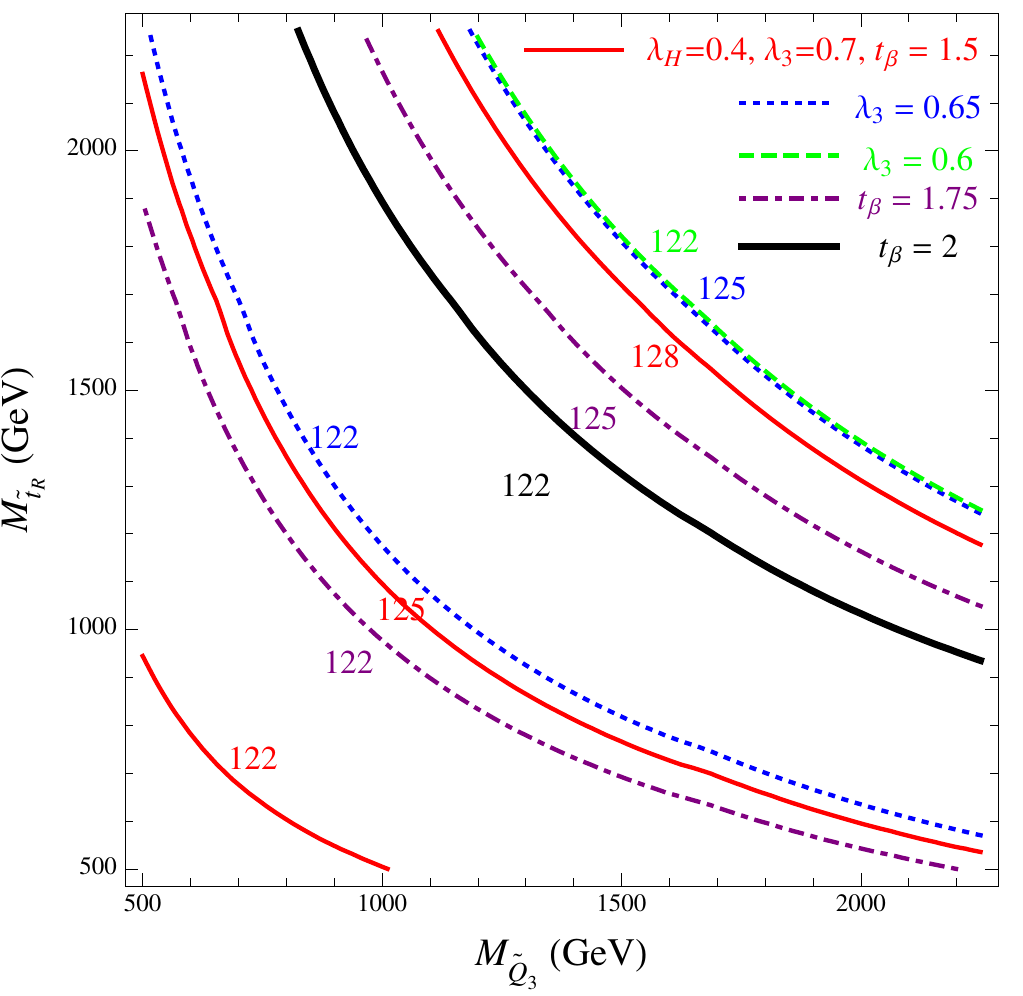}
	\caption{Lines of constant Higgs mass at the two loop order of 122, 125 and 128 GeV for various values of the parameters in the plane of the stop soft mass parameters. The red solid lines correspond to the point $\lambda_H = 0.4, \ \lambda_3 = 0.7$ and $\tan \beta = 1.5$, while subsequent lines only alter the parameter indicated on the plot, \textit{e.g.} the blue dotted line correspond to $\lambda_H = 0.4, \ \lambda_3 = 0.65$ and $\tan \beta = 1.5$. Varying $\lambda_H$ has a similar effect to varying $\lambda_3$ and so was not implemented in this plot.}
\label{img.mh.2.loop}
\end{figure}

To complete the study of the Higgs mass,  the radiatively corrected Higgs mass is displayed at the two loop level in Fig.~\ref{img.mh.2.loop}. Here, lines of constant Higgs mass are plotted in the right-handed ($M_{\tilde t_R }$) - left-handed ($M_{\tilde Q_3}$) stop soft mass parameter plane for a Higgs mass of 122, 125 and 128 GeV for different values of $\lambda_3$ and $\tan \beta$. The red solid lines correspond to the point $\lambda_H = 0.4, \ \lambda_3 = 0.7$ and $\tan \beta = 1.5$, while subsequent lines only alter the parameter indicated on the plot, \textit{e.g.} the blue dotted line correspond to $\lambda_H = 0.4, \ \lambda_3 = 0.65$ and $\tan \beta = 1.5$. Varying $\lambda_H$ has a similar effect to varying $\lambda_3$ and so was not implemented in this plot. The remainder of the relevant parameters are $(\kappa, \lambda_\Sigma) = (-0.4, 0.2)$ and $(a_H, a_3, a_\Sigma, a_\kappa, v_s, v_\Sigma) = (-10, -30, 30, 200, -1110, 1)$ GeV. The two loop contribution to the Higgs mass was calculated using the MSSM contribution from FeynHiggs~\cite{Frank:2006yh}.

\begin{figure}[h!] 
	\includegraphics[scale=0.7]{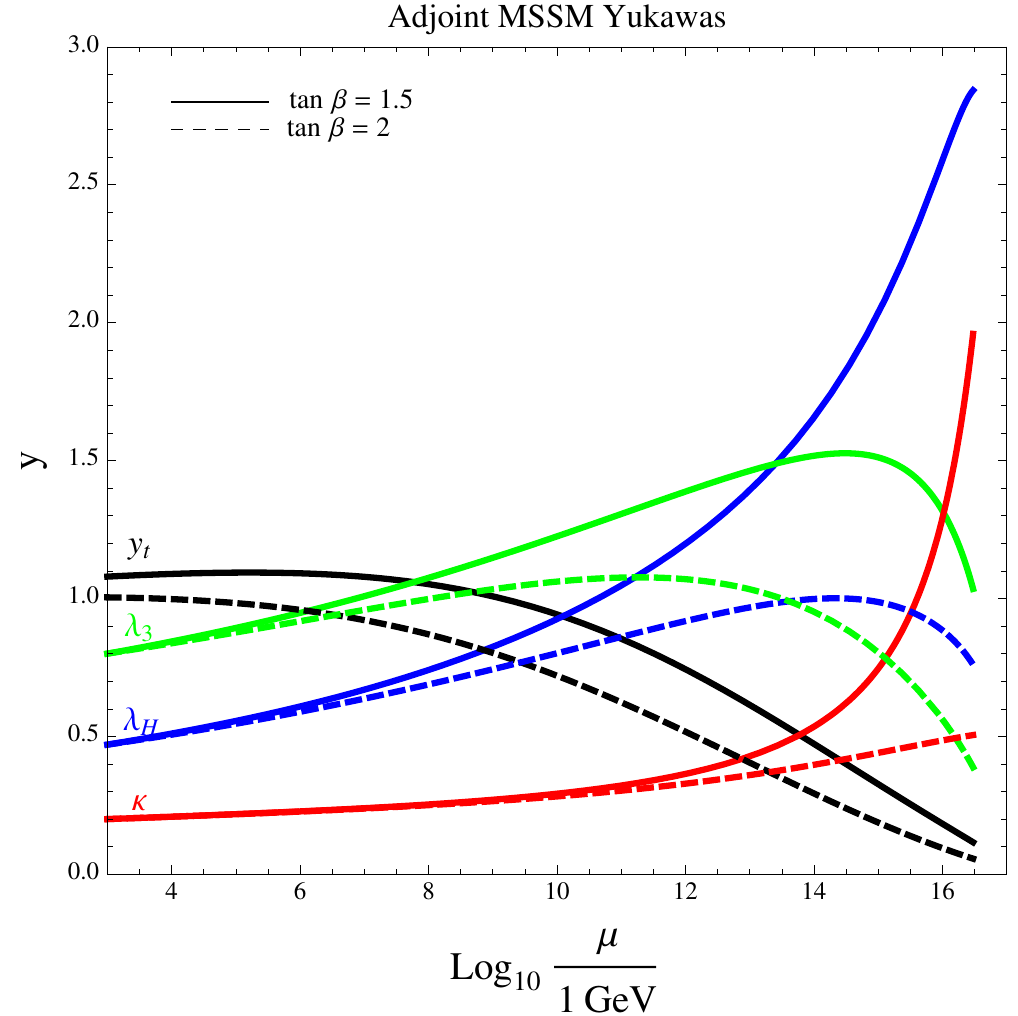}
	\caption{Evolution of the couplings in the Adjoint MSSM, where solid (dashed) curves correspond to $\tan \beta =2$ ($\tan \beta =1.5$).}
\label{img.AdjMSSM.y}
\end{figure}

In order to illustrate the behavior of the Yukawa couplings we show in Fig.~\ref{img.AdjMSSM.y} the running Yukawa couplings in the Adjoint MSSM, 
where solid (dashed) curves correspond to $\tan \beta =2$ ($\tan \beta =1.5$). Here the initial values were chosen to lie close to the non-perturbative 
regime, different than the previous plots of this nature. The evolution of the Yukawa couplings is very interesting. Due to the existence of the new colored fields the Yukawa 
coupling of the top quark can be very small at the high scale, while it is $\kappa$ and $\lambda_H$ that saturate the bounds on perturbativity. While the behavior with $\tan \beta$ has improved, decreasing it stills leads to worse behavior due to its influence on the top Yukawa coupling at the low scale.

\begin{figure}[h] 
	\includegraphics[scale=0.7]{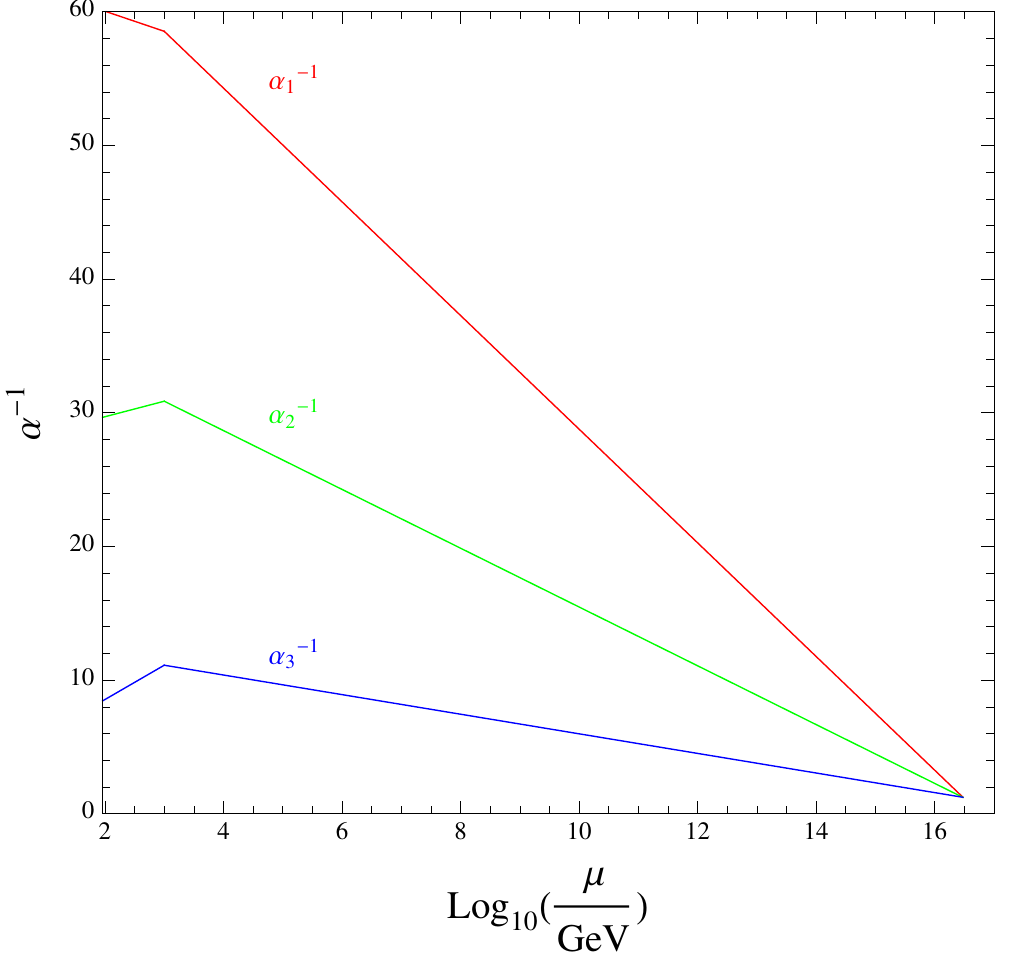}
	\caption{Evolution of the gauge couplings in the Adjoint MSSM.}
\label{img.GUT}
\end{figure}

Before we discuss the phenomenological aspects of the model we show a benchmark scenario and 
the values for the Higgs mass. Using the benchmark point given in Table~\ref{benchmark} the corresponding Higgs spectrum and its composition is shown 
in Table~\ref{spectrum}. Here we use the notation $H_i, A_i$ and $H^+_i$ for CP-even, CP-odd and charged scalars respectively, while the SM-like Higgs is denoted as $h$. In this particular point the lightest neutral Higgs is SM-like 
since the pseudo-scalar Higgs and the other Higgses are heavy, \textit{i.e.} the decoupling limit in the Higgs sector.
\subsection{Adjoint Phenomenology}
The adjoint MSSM predicts the existence of new colored fields, which acquire mass after symmetry breaking. Therefore, they have to be at the TeV scale as the upper bound on their masses is basically defined by the vacuum expectation value of the singlet field. 
The decays of the superpartners of the adjoint fields depend, of course, on the SUSY mass spectrum and the status of R-parity. 
Therefore, let us begin by talking about the new scalar fields:
\begin{itemize}
\item \underline{Colored Octet}:
The scalar octet, $\Phi \sim (8,1,0)$, can be produced with very large cross sections at the LHC and can decay 
into two gluons  at the one-loop level. Since the QCD background for thee types of channels is quite severe the 
bounds on the colored octet are not very strong. Therefore, one can have a colored octet with mass around 
500 GeV or less in agreement with all experimental constraints coming from ATLAS and CMS. 
The fermionic octet mass, $\tilde{\Phi}$, is simply
\begin{equation}
	M_{\tilde{\Phi}} = \frac{1}{\sqrt 2} \lambda_\Phi v_S.
\end{equation}
As in the case of the scalar octet, one can produce the fermionic octet in pairs through QCD interactions 
and it will decay into the scalar octet and a gluino if this channel is open. Therefore, here 
we can have channels with multi-jets and no missing energy when we include the effect of the $\lambda^{''}$ couplings.
\item \underline{Colored Doublets}:
The fields $\hat{X}^T=(\hat{X}_1, \hat{X}_2) \sim (3,2,-5/6)$ and $\hat{\bar X}^T=(\hat{\bar X}_1, \hat{\bar X}_2) \sim (\bar{3},2,5/6)$ 
can give rise to interesting phenomenology. Here the electric charge of these colored fields are $Q(X_1)=-1/3$, $Q(X_2)=-4/3$, 
$Q(\bar{X}_1)=4/3$, and $Q(\bar{X}_2)=1/3$. The mass of the fermionic fields is simply given by
\begin{equation}
	\mu_X = \frac{1}{\sqrt 2} \lambda_X v_S.
\end{equation}
and they can decay through $y_X$ into a Higgs and down quark. In the case of the scalar colored fields one can say that they 
can decay into a down quark and a neutralino, or a down quark and a chargino. Even if the production cross 
section at the LHC is large the bounds on their masses are weak due to the severe background. 
\end{itemize}
%
\section{Summary and Outlook}
We have discussed the Higgs mass in different extensions of the Minimal Supersymmetric Standard Model with the following question in mind: Is it possible for the tree-level Higgs mass to reach $110 \pm 5$ GeV with all couplings perturbative up to the GUT scale? The value of 110 GeV tree-level Higgs mass has been chosen since 500 GeV stops with little mixing are enough to raise it to the measured value of 125 GeV. Applying this guideline to both the NMSSM and the NMSSM with a real triplet yielded null results and furthermore the latter model does not lead to gauge coupling unification.

We have proposed a new extension of the MSSM where the new fields composed a complete 24 representation of $SU(5)$ and we refer to this model 
as the ``Adjoint MSSM". In this context one can generate two extra contributions to the Higgs mass at tree-level due to the existence of the 
singlet and the triplet with zero hypercharge, as in the TNMSSM. Applying a $Z_3$ symmetry to this model furthermore explains the origin of all supersymmetric mass terms (as in the NMSSM) and predicts the existence of 
light colored Higgses which could give rise to exotic and rich phenomenology at the Large Hadron Collider. Thanks to the existence 
of these extra fields with $SU(3)$ quantum numbers, $(8,1,0)$, $(3, 2, -\frac{5}{6})$,  and $(\bar 3, 2, \frac{5}{6})$, the perturbative behavior of the Yukawa couplings is greatly improved so that while there is no new contributions to the Higgs mass compared to the TNMSSM, it is possible to 
achieve a large Higgs mass in agreement with our guideline, gauge coupling unification and all relevant couplings can be perturbative up the GUT scale. 

The Higgs production at the LHC could be modified due to the presence of the colored fields, $X$ and $\bar{X}$ and their superpartners. 
At the same time the Higgs decays into two gammas also could be modify by the charged fields in the real triplet superfield and the 
colored fields with hypercharge. See Refs.~\cite{FileviezPerez:2008bj} and~\cite{DiChiara:2008rg,Delgado:2012sm} 
for a previous study of the impact of the Higgs triplet in the radiative Higgs decays. Of course, the new colored fields can be 
around the TeV scale and their impact on the production and decays of the Higgs could be insignificant. 
These and other aspects of the theory, including a possible UV completion, will be investigated in a future publication. 

\begin{widetext}
\appendix
{\bf{Appendix A: CP-even Neutral Higgs Masses}}

In this appendix we show the mass of the neutral CP-even Higgses neglecting 
the terms proportional to $v_{\Sigma}^2$. In the basis 
$\sqrt{2} \text{Re}(H_d^0,H_u^0, S, \Sigma^0)$, the CP-even scalar mass matrix is
\begin{eqnarray}
	\left(M_S^2\right)_{11} & = & c_\beta^2 M_Z^2
	 - t_\beta \left(\frac{1}{\sqrt 2} a_H v_S + \frac{1}{2} \lambda_H \kappa  v_S^2\right)
	- \frac{1}{2} t_\beta v_\Sigma \left(a_3 + \frac{1}{\sqrt 2} \lambda_3 \lambda_\Sigma v_S \right),
	\\
	\left(M_S^2\right)_{12} & = & \frac{1}{\sqrt 2} a_H v_S + \frac{1}{2} \lambda_H \kappa v_S^2+ \frac{1}{4} s_{2 \beta}
	\left(
		2 \lambda_H^2 v^2 +  \lambda_3^2 v^2 - 2 M_Z^2
	\right)
	 \ + \  \frac{1}{2} v_\Sigma \left(a_3 + \frac{1}{\sqrt 2} \lambda_3 \lambda_\Sigma v_S\right)
\\
	\left(M_S^2\right)_{13} & = & v
	\left[
		c_\beta \lambda_H^2 v_S + \frac{1}{\sqrt 2} s_\beta \, a_H  + s_\beta \, \lambda_H \kappa v_S
		+ \frac{1}{\sqrt 2} \lambda_3 v_\Sigma
		\left(
			c_\beta \lambda_H + \frac{1}{2} s_\beta \lambda_\Sigma
		\right)
	\right],
\\
	\left(M_S^2\right)_{14} & = & v
	\left[
		\frac{1}{\sqrt 2} c_\beta \lambda_3 \lambda_H v_S + \frac{1}{2} s_\beta a_3
		+ \frac{1}{2 \sqrt 2} s_\beta \lambda_3 \lambda_\Sigma v_S
		+ \frac{1}{2}v_\Sigma
		\left(
			c_\beta \lambda_3^2 + s_\beta \lambda_H \lambda_\Sigma
		\right)
	\right],
\\
	\left(M_S^2\right)_{22} & = & s_\beta^2 M_Z^2
	 - \frac{1}{t_\beta} \left(\frac{1}{\sqrt 2} a_H v_S + \frac{1}{2} \lambda_H \kappa  v_S^2\right)
	- \frac{1}{2 t_\beta}  v_\Sigma \left(a_3 + \frac{1}{\sqrt 2} \lambda_3 \lambda_\Sigma v_S \right),
\\
	\left(M_S^2\right)_{23} & = &v
	\left[
		s_\beta \lambda_H^2 v_S + \frac{1}{\sqrt 2} c_\beta \, a_H  + c_\beta \, \lambda_H \kappa v_S
		+ \frac{1}{\sqrt 2} \lambda_3 v_\Sigma
		\left(
			s_\beta \lambda_H + \frac{1}{2} c_\beta \lambda_\Sigma
		\right)
	\right],
\\
	\left(M_S^2\right)_{24} & = & v
	\left[
		\frac{1}{\sqrt 2} s_\beta \lambda_3 \lambda_H v_S + \frac{1}{2} c_\beta a_3
		+ \frac{1}{2 \sqrt 2} c_\beta \lambda_3 \lambda_\Sigma v_S
		+ \frac{1}{2}v_\Sigma
		\left(
			s_\beta \lambda_3^2 + c_\beta \lambda_H \lambda_\Sigma
		\right)
	\right],
\\
	\left(M_S^2\right)_{33} & = & 2 \kappa^2 v_S^2 + \frac{1}{\sqrt 2} a_\kappa v_S - \frac{1}{2 \sqrt 2} s_{2 \beta} \frac{a_H v^2}{v_S}
	- \frac{1}{4 \sqrt 2} \lambda_3  \frac{v^2}{v_S}v_\Sigma
	\left(
		s_{2\beta} \lambda_\Sigma + 2 \lambda_H
	\right),
\\
	\left(M_S^2\right)_{34} & = & 
		\frac{1}{4 \sqrt 2 } \lambda_3 v^2
		\left(
			2 \lambda_H + s_{2 \beta} \lambda_\Sigma
		\right)
		+ v_\Sigma
		\left(
			\frac{1}{\sqrt 2} a_\Sigma + \kappa \lambda_\Sigma v_S+ \lambda_\Sigma^2 v_S
		\right),
\\
	\left(M_S^2\right)_{44} & = & -\frac{v^2}{4 \sqrt 2 v_\Sigma}
	\left[
		s_{2 \beta} \left(\sqrt 2 a_3 + \lambda_3 \lambda_\Sigma v_S\right)
		+ 2 \lambda_3 \lambda_H v_S
	\right].
\end{eqnarray}

{\bf{Appendix B: CP-odd Neutral Higgs Masses:}}
In the basis $\sqrt{2} \text{Im}(H_d^0,H_u^0, S, \Sigma^0)$, the CP-odd scalar mass matrix is given by
\begin{eqnarray}
	\left(M_P^2\right)_{11} & = & -t_\beta v_S
	\left(
		\frac{1}{\sqrt 2} a_H + \frac{1}{2} \lambda_H \kappa v_S
	\right)
	- \frac{1}{2} t_\beta v_\Sigma
	\left(
		a_3 + \frac{1}{\sqrt 2}\lambda_3 \lambda_\Sigma v_S
	\right),
\\
	\left(M_P^2\right)_{12} & = & -\frac{1}{\sqrt 2} a_H v_S -\frac{1}{2} \lambda_H \kappa v_S^2
	- \frac{1}{2} v_\Sigma
	\left(
		a_3 + \frac{1}{\sqrt 2} \lambda_3 \lambda_\Sigma v_S
	\right),
\\
%
	\left(M_P^2\right)_{13} & = & s_\beta v
	\left(
		-\frac{1}{\sqrt 2} a_H + \lambda_H \kappa v_S + \frac{1}{2 \sqrt 2}v_\Sigma  \lambda_3 \lambda_\Sigma
	\right),
\\
	\left(M_P^2\right)_{14} & = & \frac{1}{2} s_\beta v
	\left(
		-a_3 + \frac{1}{\sqrt 2} \lambda_3 \lambda_\Sigma v_S  + v_\Sigma \lambda_H \lambda_\Sigma
	\right),
\\
	\left(M_P^2\right)_{22} & = & -\frac{v_S}{t_\beta} \left(\frac{1}{\sqrt 2} a_H  + \frac{1}{2} \lambda_H \kappa v_S \right)
	- \frac{1}{2 \, t_\beta}v_\Sigma
	\left(
		a_3 + \frac{1}{\sqrt 2}\lambda_3 \lambda_\Sigma v_S
	\right),
\\
	\left(M_P^2\right)_{23} & = & c_\beta v
	\left(
		-\frac{1}{\sqrt 2} a_H + \lambda_H \kappa v_S + \frac{1}{2 \sqrt 2} v_\Sigma \lambda_3 \lambda_\Sigma
	\right),
\\
	\left(M_P^2\right)_{24} & = &
	\frac{1}{2} c_\beta v
	\left(
		-a_3 + \frac{1}{\sqrt 2} \lambda_3 \lambda_\Sigma v_S + v_\Sigma \lambda_H \lambda_\Sigma
	\right),
\\
	\left(M_P^2\right)_{33} & =  & - \frac{3}{\sqrt 2} a_\kappa v_S - \frac{1}{2} s_{2 \beta} \frac{v^2}{v_S} \left(\frac{1}{\sqrt 2}a_H + 2 \lambda_H \kappa v_S\right)
	- \frac{1}{4 \sqrt 2} \lambda_3 \frac{v^2}{v_S} v_\Sigma
	\left(
		s_{2\beta} \lambda_\Sigma + 2 \lambda_H
	\right) 
\\
	\left(M_P^2\right)_{34} & = & 
		\frac{1}{4 \sqrt 2} \lambda_3 v^2
		\left(
			2 \lambda_3 \lambda_H - \frac{1}{\sqrt 2} s_{2 \beta} \lambda_\Sigma
		\right)
		+ v_\Sigma
		\left(
			\lambda_\Sigma \kappa v_S - \frac{1}{\sqrt 2} a_\Sigma
		\right),
\\
	\left(M_P^2\right)_{44} & = & - \sqrt 2 a_\Sigma v_S - \lambda_\Sigma \kappa v_S^2  - \frac{1}{2} s_{2 \beta} \lambda_H \lambda_\Sigma v^2
	-\frac{v^2}{4v_\Sigma} 
	\left[
		s_{2 \beta}
		\left(
			a_3 + \frac{1}{\sqrt 2}\lambda_3 \lambda_\Sigma v_S
		\right)
		+ \sqrt 2 \lambda_3 \lambda_H v_S
	\right].
	\nonumber \\
\end{eqnarray}
where of course the determinant is zero.

{\bf{Appendix C: Charged Higgs Masses}}

In the basis $\left(H_d^-, H_u^{+*}, \Sigma^-, \Sigma^{+*}\right)$, the charged scalars mass matrix is
\begin{eqnarray}
	\left(M_{\pm}^2\right)_{11} & = &  s_\beta^2
	\left(
		M_W^2 + \frac{1}{4}\lambda_3^2 v^2 - \frac{1}{2 }\lambda_H^2 v^2
	\right)
	- t_\beta \left(\frac{1}{\sqrt 2} a_H v_S + \frac{1}{2} \lambda_H \kappa v_S^2\right)
\nonumber 	\\
	& & 
	- v_\Sigma
	\left[
		\frac{1}{2} t_\beta
		\left(
			a_3 + \frac{1}{\sqrt 2}\lambda_3 \lambda_\Sigma v_S
		\right)
		+\sqrt 2 \lambda_3 \lambda_H v_S
	\right],
\\
	\left(M_{\pm}^2\right)_{12} & = & -\frac{1}{\sqrt 2} a_H v_S - \frac{1}{2} \lambda_H \kappa v_S^2+ \frac{1}{2} s_{2 \beta}
	\left(
		M_W^2 + \frac{1}{4} \lambda_3^2 v^2 - \frac{1}{2}\lambda^2 v^2
	\right)
	+\frac{1}{2}v_\Sigma \left( a_3 + \frac{1}{\sqrt 2 }\lambda_3 \lambda_\Sigma v_S \right),
	\nonumber \\
\\
	\left(M_{\pm}^2\right)_{13} & = & \frac{1}{2} \lambda_3 v v_S \left( c_\beta \lambda_H + s_\beta \lambda_\Sigma \right)
	+ \sqrt 2 \, c_\beta \frac{v_\Sigma}{v}
	\left(
		M_W^2 - \frac{1}{4} \lambda_3^2 v^2
	\right),
\\
	\left(M_{\pm}^2\right)_{14} & = & \frac{1}{\sqrt 2} v \left(\frac{1}{\sqrt 2}c_\beta \lambda_H \lambda_3 v_S + s_\beta a_3 \right)
	-\sqrt 2 \, c_\beta \frac{v_\Sigma}{v}
	\left(
		M_W^2 - \frac{1}{4} \lambda_3^2 v^2
	\right),
\\
	\left(M_{\pm}^2\right)_{22} & =  & c_\beta^2
	\left(
		M_W^2 + \frac{1}{4}\lambda_3^2 v^2 - \frac{1}{2 }\lambda_H^2 v^2
	\right)
	- \frac{1}{t_\beta} \left(\frac{1}{\sqrt 2} a_H v_S + \frac{1}{2} \lambda_H \kappa v_S^2\right)
\nonumber 	\\
	& &-v_\Sigma
	\left[
		\frac{1}{2 t_\beta }
		\left(
			a_3 + \frac{1}{\sqrt 2}\lambda_3 \lambda_\Sigma v_S
		\right)
		+\sqrt 2 \lambda_3 \lambda_H v_S
	\right],
\\
	\left(M_{\pm}^2\right)_{23} & = & - v
	\left(
		\frac{1}{\sqrt 2} c_\beta a_3 + \frac{1}{2} s_\beta \lambda_H \lambda_3 v_S
	\right)
	+ \sqrt 2 s_\beta \frac{v_\Sigma}{v}
	\left(
		M_W^2 - \frac{1}{4} \lambda_3^2 v^2
	\right),
\\
	\left(M_{\pm}^2\right)_{24} & = & -\frac{1}{2} \lambda_3 v v_S \left( s_\beta \lambda_H + c_\beta \lambda_\Sigma \right)
	- \sqrt 2 \, s_\beta \frac{v_\Sigma}{v}
	\left(
		M_W^2 - \frac{1}{4} \lambda_3^2 v^2
	\right),
\end{eqnarray}
\begin{eqnarray}
	\left(M_{\pm}^2\right)_{33} & = & -\frac{1}{\sqrt 2} a_\Sigma v_S - \frac{1}{2} \kappa \lambda_\Sigma v_S^2 - c_{2 \beta} \left(M_W^2 - \frac{1}{4} \lambda_3^2 v^2\right)
	- \frac{1}{4} s_{2 \beta} \lambda_H \lambda_\Sigma v^2
\nonumber \\
	& &
	- \frac{v^2}{4v_\Sigma}
	\left[
		s_{2 \beta}
		\left(
			a_3 + \frac{1}{\sqrt 2} \lambda_3 \lambda_\Sigma v_S
		\right)
		+ \sqrt 2 \lambda_3 \lambda_H v_S
	\right],
\\
	\left(M_{\pm}^2\right)_{34} & = & \frac{1}{\sqrt 2} a_\Sigma v_S
	+ \frac{1}{2} \lambda_\Sigma \kappa v_S^2
	+ \frac{1}{4} s_{2\beta} \lambda_H \lambda_\Sigma v^2,
\\
	\left(M_{\pm}^2\right)_{44} & = & -\frac{1}{\sqrt 2} a_\Sigma v_S - \frac{1}{2} \lambda_\Sigma \kappa v_S^2 + c_{2 \beta} \left(M_W^2 - \frac{1}{4} \lambda_3^2 v^2\right)
	- \frac{1}{4} s_{2 \beta} \lambda_H \lambda_\Sigma v^2
\nonumber \\
	& &
	- \frac{v^2}{4v_\Sigma}
	\left[
		s_{2 \beta}
		\left(
			a_3 + \frac{1}{\sqrt 2} \lambda_3 \lambda_\Sigma v_S
		\right)
		+ \sqrt 2 \lambda_3 \lambda_H v_S
	\right].
\end{eqnarray}
{\bf{Appendix D: Fermionic Spectrum:}}
In the basis $\left(\tilde W^- \tilde H_d^-, \tilde \Sigma^-\right)$ by $\left(\tilde W^+ \tilde H_u^+, \tilde \Sigma^+\right)$ 
the chargino mass matrix has block off-diagonal entries, $X$ and $X^T$ with
\begin{equation}
	X =
	\begin{pmatrix}
		M_2
		&
		\sqrt 2 s_\beta M_W
		&
		-2 \frac{v_\Sigma}{v} M_W
		\\
		\sqrt 2 c_\beta M_W
		&
		\frac{1}{2} \lambda_3 v_\Sigma - \frac{1}{\sqrt 2} \lambda_H v_S
		&
		\frac{1}{\sqrt 2} s_\beta \lambda_3 v
		\\
		2 \frac{v_\Sigma}{v} M_W
		&
		-\frac{1}{\sqrt 2} c_\beta \lambda_3 v
		&
		\frac{1}{\sqrt 2} \lambda_\Sigma v_S
	\end{pmatrix}.
\end{equation}
In the basis $\left(\tilde B, \tilde W^0, \tilde H_d^0, \tilde H_u^0, \tilde S, \tilde \Sigma^0\right)$ the neutralino mass matrix is
\begin{equation}
	M_{\chi^0} =
	\begin{pmatrix}
		M_1
		&
		0
		&
		-s_W c_\beta M_Z
		&
		s_W s_\beta M_Z
		&
		0
		&
		0
		\\
		0
		&
		M_2
		&
		c_W c_\beta M_Z
		&
		-c_W s_\beta M_Z
		&
		0
		&
		0
		\\
		- s_W c_\beta M_Z
		&
		c_W c_\beta M_Z
		&
		0
		&
		\frac{1}{2} \lambda_3 v_\Sigma + \frac{1}{\sqrt 2} \lambda_H v_S
		&
		\frac{1}{\sqrt 2 } s_\beta \lambda_H v
		&
		\frac{1}{ 2 } s_\beta \lambda_3 v
		\\
		s_W s_\beta M_Z
		&
		- c_W s_\beta M_Z
		&
		\frac{1}{2} \lambda_3 v_\Sigma + \frac{1}{\sqrt 2} \lambda_H v_S
		&
		0
		&
		\frac{1}{\sqrt 2} c_\beta \lambda_H v
		&
		\frac{1}{2} c_\beta \lambda_3 v
		\\
		0
		&
		0
		&
		\frac{1}{\sqrt 2 } s_\beta \lambda_H v
		&
		\frac{1}{\sqrt 2} c_\beta \lambda_H v
		&
		\sqrt 2 \kappa v_S
		&
		\frac{1}{\sqrt 2} \lambda_\Sigma v_\Sigma
		\\
		0
		&
		0
		&
		\frac{1}{2} s_\beta \lambda_3 v
		&
		\frac{1}{2} c_\beta \lambda_3 v
		&
		\frac{1}{\sqrt 2} \lambda_\Sigma v_\Sigma
		&
		\frac{1}{\sqrt 2} \lambda_\Sigma v_S
	\end{pmatrix}.
\end{equation}
{\bf{Appendix E: Renormalization Group Equations}}:
The equations which describe the evolution of the gauge couplings is given by 
\begin{eqnarray}
	\alpha_1(M_\text{GUT})^{-1} & = & \alpha_1^{-1}(M_Z) - \frac{1}{2 \pi}
	\left(
		\frac{41}{10} \log \frac{M_\text{GUT}}{M_Z}  + \frac{4}{3} \log \frac{M_\text{GUT}}{M_{\tilde f_{1,2}}}
		+ \frac{2}{3} \log \frac{M_\text{GUT}}{M_{\tilde f_{3}}}+ \frac{1}{10} \log \frac{M_\text{GUT}}{M_A}
	\right)
\nonumber \\
	& & 
		- \frac{1}{2 \pi} \left(
		  \frac{2}{5} \log \frac{M_\text{GUT}}{M_{\tilde H_{u,d}}}
		  + 5 \log \frac{M_\text{GUT}}{M_X}
	\right),
	\\
	\alpha_2(M_\text{GUT})^{-1} & = & \alpha_2^{-1}(M_Z) - \frac{1}{2 \pi}
	\left(
		-\frac{19}{6} \log \frac{M_\text{GUT}}{M_Z} + \frac{4}{3} \log \frac{M_\text{GUT}}{M_{\tilde W}}
		 + \frac{4}{3} \log \frac{M_\text{GUT}}{M_{\tilde f_{1,2}}}+ \frac{2}{3} \log \frac{M_\text{GUT}}{M_{\tilde f_{3}}}
	\right.
\nonumber	\\
	& & \left. \quad \quad \quad \quad \quad \quad  \quad \ \ \
		 + \frac{1}{6} \log \frac{M_\text{GUT}}{M_A} + \frac{2}{3} \log \frac{M_\text{GUT}}{M_{\tilde H_{u,d}}}
		 + 2 \log \frac{M_\text{GUT}}{M_\Sigma} + 3 \log \frac{M_\text{GUT}}{M_X}
	\right),
	 \\
	\alpha_3(M_\text{GUT})^{-1} & = & \alpha_3^{-1}(M_Z) - \frac{1}{2 \pi}
	\left(
		-7 \log \frac{M_\text{GUT}}{M_Z} + 2 \log \frac{M_\text{GUT}}{M_{\tilde g}}
		 + \frac{4}{3} \log \frac{M_\text{GUT}}{M_{\tilde f_{1,2}}}+ \frac{2}{3} \log \frac{M_\text{GUT}}{M_{\tilde f_{3}}}
	\right)
\nonumber	\\
	& & 
	 - \frac{1}{2 \pi} \left( 3 \log \frac{M_\text{GUT}}{M_\Phi} + 2 \log \frac{M_\text{GUT}}{M_X}
	\right).
\end{eqnarray}
assuming the threshold of the new particles as well as the SUSY fields is at one TeV. The running is shown in Fig~\ref{img.GUT}.

The RGE for a given Yukawa coupling is simply the sum of the anomalous dimensions of the all three particles involved in
 that coupling multiplied by that Yukawa coupling:
\begin{eqnarray}
	\label{beta.y.top}
	16 \pi^2 \frac{d}{dt} y_t & = & y_t
	\left(
		6 y_t^2 + y_b^2 + 3 y_X^2 + \frac{3}{2} \lambda_3^2 + \lambda_H^2
		- \frac{16}{3} g_3^2 - 3 g_2^2 - \frac{13}{15} g_1^2
	\right),
\end{eqnarray}
\begin{eqnarray}
	16 \pi^2 \frac{d}{dt} y_b & = & y_b
	\left(
		6y_b^2 + y_t^2 +y_\tau^2 + 2 y_X^2 + \frac{3}{2} \lambda_3^2 + \lambda_H^2
		- \frac{16}{3} g_3^2 - 3 g_2^2 - \frac{7}{15} g_1^2
	\right),
	\\
	16 \pi^2 \frac{d}{dt} y_\tau & = & y_\tau
	\left(
		4 y_\tau^2 + 3 y_b^2  + \frac{3}{2} \lambda_3^2 + \lambda_H^2 -  3 g_2^2 - \frac{9}{5} g_1^2
	\right),
	\\
	16 \pi^2 \frac{d}{dt} y_X & = & y_X
	\left(
		6 y_X^2+3 y_t^2+2 y_b^2 + \lambda_X^2 + \frac{3}{2} \lambda_3^2 + \lambda_H^2 + \frac{16}{3} \zeta_\Phi^2
		+ \frac{3}{2} \zeta_\Sigma^2  - \frac{16}{3} g_3^2- 3 g_2^2 - \frac{19}{15} g_1^2
	\right),
\\
	16 \pi^2 \frac{d}{dt} \eta & = & \eta
	\left(
		40 \eta^2 + 3 \lambda_\Phi^2 + 6 \zeta_\Phi^2- 18 g_3^2
	\right),
\\
	16 \pi^2 \frac{d}{dt} \lambda_X & = & \lambda_X
	\left(
		8 \lambda_X^2 + y_X^2+ 4 \lambda_\Phi^2 + 2 \lambda_H^2 + \frac{3}{2} \lambda_\Sigma^2+2 \kappa^2
		+ \frac{32}{3} \zeta_\Phi^2 + 3 \zeta_\Sigma^2
		  - \frac{16}{3} g_3^2 - 3 g_2^2 - \frac{5}{3} g_1^2
	\right),
	\\
	16 \pi^2 \frac{d}{dt} \lambda_\Phi & = & \lambda_\Phi
	\left(
		6 \lambda_\Phi^2 + \frac{80}{3} \eta^2+6 \lambda_X^2 + 2 \lambda_H^2 + \frac{3}{2} \lambda_\Sigma^2+2 \kappa^2
		+4 \zeta_\Phi^2 - 12 g_3^2
	\right),
\\
	16 \pi^2 \frac{d}{dt} \lambda_3 & = & \lambda_3
	\left(
		4 \lambda_3^2 + 3 y_t^2 +3 y_b^2 + y_\tau^2+ 3 y_X^2 + 2\lambda_H^2+ \lambda_\Sigma^2
		+ 3 \zeta_\Sigma^2 -7 g_2^2 - \frac{3}{5} g_1^2
	\right),
	\\
	16 \pi^2 \frac{d}{dt} \lambda_H & = & \lambda_H
	\left(
		4 \lambda_H^2 + 3 y_t^2 + 3 y_b^2 + y_\tau^2 + 3 y_X^2 + 6 \lambda_X^2
		 + 4 \lambda_\Phi^2+3 \lambda_3^2 + \frac{3}{2} \lambda_\Sigma^2+2 \kappa^2
		 -3 g_2^2 - \frac{3}{5} g_1^2
	\right),
	\\
	16 \pi^2 \frac{d}{dt} \lambda_\Sigma & = & \lambda_\Sigma
	\left(
		\frac{7}{2} \lambda_\Sigma^2 + 6 \lambda_X^2 + 4 \lambda_\Phi^2 + 2 \lambda_3^2 + 2 \lambda_H^2 + 2 \kappa^2
		+6\zeta_\Sigma^2- 8 g_2^2
	\right),
	\\
	16 \pi^2 \frac{d}{dt} \kappa & = & \kappa
	\left(
		6 \kappa^2 + 18 \lambda_X^2 + 12 \lambda_\Phi^2 + 6 \lambda_H^2 + \frac{9}{2} \lambda_\Sigma^2
	\right),
	\\
	16 \pi^2 \frac{d}{dt} \zeta_\Phi & = & \zeta_\Phi
	\left(
		y_X^2 + 2 \lambda_X^2 +  \frac{40}{3} \eta^2 + \lambda_\Phi^2 + \frac{38}{3} \zeta_\Phi^2
	+ 3 \zeta_\Sigma^2- \frac{34}{3} g_3^2 - 3 g_2^2 - \frac{5}{3} g_1^2
	\right),
	\\
	16 \pi^2 \frac{d}{dt} \zeta_\Sigma & = & \zeta_\Sigma
	\left(
		y_X^2 + 2 \lambda_X^2 +  \lambda_3^2 + \lambda_\Sigma^2  + \frac{32}{3} \zeta_\Phi^2
	+ 6 \zeta_\Sigma^2 - \frac{16}{3} g_3^2 - 7 g_2^2 - \frac{5}{3} g_1^2
	\right).
\end{eqnarray}
The anomalous dimensions of the particle content are
\begin{align}
	16 \pi^2 \gamma_{Q_3} &= y_t^2 + y_b^2 - \frac{8}{3} g_3^2 - \frac{3}{2} g_2^2 - \frac{1}{30} g_1^2,
	\\
	16 \pi^2 \gamma_{t^c} &= 2 y_t^2 - \frac{8}{3} g_3^2 - \frac{8}{15} g_1^2,
	\\
	16 \pi^2 \gamma_{b^c} &= 2 y_b^2 + 2 y_X^2- \frac{8}{3} g_3^2 - \frac{2}{15} g_1^2,
	\\
	16 \pi^2 \gamma_{L_3} &= y_\tau^2 -  \frac{3}{2} g_2^2 - \frac{3}{10} g_1^2,
	\\
	16 \pi^2 \gamma_{\tau^c} &= 2 y_\tau^2 - \frac{6}{5} g_1^2,
	\\
	16 \pi^2 \gamma_{X} &= y_X^2 + \lambda_X^2 + \frac{16}{3} \zeta_\Phi^2
	+ \frac{3}{2} \zeta_\Sigma^2- \frac{8}{3} g_3^2 - \frac{3}{2} g_2^2 - \frac{5}{6} g_1^2,
	\\
	16 \pi^2 \gamma_{\bar X} &= \lambda_{X}^2 + \frac{16}{3} \zeta_\Phi^2
	+ \frac{3}{2} \zeta_\Sigma^2 - \frac{8}{3} g_3^2 - \frac{3}{2} g_2^2 - \frac{5}{6} g_1^2,
	\\
	16 \pi^2 \gamma_{\Phi} &= \frac{40}{3} \eta^2 + \lambda_\Phi^2 + 2 \zeta_\Phi^2- 6 g_3^2,
	\\
	16 \pi^2 \gamma_{S} &= 6 \lambda_X^2 + 4 \lambda_\Phi^2 + 2 \lambda_H^2 + \frac{3}{2} \lambda_\Sigma^2+2 \kappa^2,
	\\
	16 \pi^2 \gamma_{H_u} &= 3 y_t^2 + 3 y_X^2 + \frac{3}{2} \lambda_3^2 + \lambda_H^2 -  \frac{3}{2} g_2^2 - \frac{3}{10} g_1^2,
\\
	16 \pi^2 \gamma_{H_d} &= 3 y_b^2 + y_\tau^2 + \frac{3}{2} \lambda_3^2 + \lambda_H^2 -  \frac{3}{2} g_2^2 - \frac{3}{10} g_1^2,
	\\
		16 \pi^2 \gamma_{\Sigma} &= \lambda_3^2 + \lambda_\Sigma^2 + 3 \zeta_\Sigma^2 - 4 g_2^2.
\end{align}

\begin{table}[h]
\begin{center}
\begin{tabular}{|c|c|c|c|c|c|c|c|c|c|c|}
	\hline
	$\tan \beta$ & $ \lambda_H $ & $\kappa$  & $\lambda_3$ & $\lambda_\Sigma$ & $v_S$ & $a_H$ & $a_\kappa$
	& $a_3$ & $a_\Sigma$ & $m_\Sigma$
	\\
	\hline
	2 & -0.4 & 0.4 & 0.8 & 0.4 & -1550 GeV & 110 GeV & 100 GeV & 90 GeV & 70 GeV & 2230 GeV
	\\
	\hline
\end{tabular}
\end{center}
\caption{Higgs sector parameters for the benchmark point. The corresponding Higgs spectrum is displayed in Table~II.}
\label{benchmark}
\end{table}
\begin{table}[h]
\begin{center}
\begin{tabular}{|c|c|c|c|c|c|c|c|c|c|c|}
	\hline
	Field & h & $H_1$ & $H_2$ & $H_3$ & $A_1$ & $A_2$ & $A_3$ & $H^+_1$ & $H^+_2$ & $H^+_3$
	\\
	\hline
	Mass (GeV )& 109 & 800 & 890 & 2730 & 580 & 890 & 2690 & 890 & 2690 & 2730
	\\
	\hline
	\% of $H_u$ & 80\% & 2\% & 17\% & 0\% & 0\% & 79\% & 0\% & 80\% & 0\% & 0\%
	\\
	\% of $H_d$ & 20\% & 9\% & 71\% & 0\% & 0\% & 20\% & 0\% & 20\% & 0\% & 0\%
	\\
	\% of $S$ & 0\% & 89\% & 11\% & 0\% & 100\% & 1\% & 0\% & 0\% & - & - 
	\\
	\% of $\Sigma$ & 0\% & 0\% & 0\% & 100\% & 0\% & 0\% & 100\% & 0\% & 51\%, 49\%& 49\%, 51\%
	\\
	\hline
\end{tabular}
\end{center}
\caption{Tree-level Higgs spectrum based on the benchmark point in Table~\ref{benchmark}. Here, $H_i, A_i$ and $H^+_i$ represent CP-even, CP-odd and charged scalars respectively labeled from lightest to heaviest in ascending order, while the SM-like Higgs is denoted as $h$. Also shown is the composition of each of the physical states in terms of the gauge states. The two different values for the $\Sigma$ composition of the heaviest charged fields is for $\Sigma^-$ and $\Sigma^+$ respectively.}
\label{spectrum}
\end{table}
\end{widetext}

\end{document}